\documentclass[letterpaper, journal]{IEEEtran}  

\usepackage{graphics} 
\usepackage{amsmath} 
\usepackage{amssymb}  
\usepackage[dvipdfmx]{graphicx}
\usepackage{booktabs}
\usepackage{cite}
\usepackage{enumerate}
\usepackage{cleveref}
\usepackage{url}
\usepackage{titlesec}
\usepackage[subrefformat=parens]{subcaption}
\newcommand{\sign}{\operatorname{sign}}
\newtheorem{theo}{Theorem}
\newtheorem{assump}{Assumption}
\newtheorem{lem}{Lemma}
\newtheorem{prop}{Proposition}
\newenvironment{proof}{\textit{Proof:}}{\hfill  $\blacksquare$}

\begin{document}

\title{Quantized Zero Dynamics Attacks against\\ Sampled-data Control Systems}

\author{Kosuke Kimura and Hideaki Ishii,~\IEEEmembership{Fellow,~IEEE}
\thanks{This work was supported in the part by JSPS under Grants-in-Aid for 
	Scientific Research Grant No.~18H01460 and 22H01508.}
\thanks{%
K.~Kimura is with the Department of 
System Control and H.~Ishii is with the
Department of Computer Science, both at
the Tokyo Institute of Technology, 
Yokohama, 226-8502, Japan. 
(e-mail: kimura@sc.dis.titech.ac.jp, ishii@c.titech.ac.jp).}}

\maketitle

\begin{abstract}
For networked control systems, cyber-security issues have gained much attention in recent years. In this paper, we consider the so-called zero dynamics attacks, which 
form an important class of false data injection attacks, with a special focus on the effects of quantization in a sampled-data control setting. When the attack signals must be quantized, some error will be necessarily introduced, potentially increasing the chance of detection through the output of the system. In this paper, we show however that the attacker may reduce such errors by avoiding to directly quantize the attack signal. We look at two approaches for generating quantized attacks which can keep the error in the output smaller than a specified level by using the knowledge of the system dynamics. The methods are based on a dynamic quantization technique 
and a modified version of zero dynamics attacks. Numerical examples are provided to verify the effectiveness of the proposed methods.
\end{abstract}

\section{Introduction}\label{ch:introduction}

Recent advances in communication technology are
enabling various cyber-physical systems (CPSs)
to be further connected by networks, leading to 
enhancements in efficiency and flexibility for
their operation and control. Application domains
where such changes have brought significant 
progresses include large-scale plants, smart grids, 
and traffic networks. It is however inevitable that 
along with the increase in network connectivity,
cyber-security issues have gained much more attention. 
Some of the major incidents include 
the attacks against nuclear research facilities in Iran
\cite{Farwell2011StuxnetAT}, 
power grids in Ukraine \cite{ukraine}, and 
sewage systems in Australia \cite{Slay2007LessonsLF}.

To maintain the defense in depth for security and safety
of such CPSs, it is of critical
importance to strengthen security methods from the
perspective of systems control
to complement conventional security methods based on 
information technologies \cite{FerTei:22,IshZhu:22}.
In particular, malicious cyber attacks against CPSs
may have physical consequences, which can 
potentially result in damages in control devices and
facilities. 

In the systems control area, recent research has studied
analyses on impacts of such attacks,
detection techniques, as well as resilient control methods.
Representative classes of attacks that may lead to manipulating
vulnerable physical plants include those of replay attacks, 
Denial-of-Service (DoS) attacks, and 
false-data injection (FDI) attacks;
see, for example, \cite{MoKimBra:12,contsec,TeiShaSan:15}
and the references therein.

The focus of this paper is on the class of FDI attacks
known as the zero dynamics attacks \cite{ParShiLee:19,TeiShaSan:15}. 
An attacker who is aware of the system dynamics may
modify the control signals in such a way that the
internal states of the system are manipulated by the
attacker and can diverge. 
They can be generated by taking advantage of 
system zeros and, in particular, the unstable ones. 
The difficulty in dealing with such attacks is that
the behavior of the system output remains the same 
as that without the attacks. Hence, it is hard to detect 
them, e.g.,
by conventional fault detection techniques.

In this paper, we formulate a networked control 
problem in the sampled-data setting, where 
the continuous-time plant is controlled by a digital 
compensator and the control 
input is transmitted over a network channel \cite{CheFra:95}.
We place particular attention to the role played by 
quantization in the control signals. 
Since zero dynamics
attacks normally require the attack signals to be continuous,
discretizing them through quantization will introduce certain errors
not only in the attacks but also in the system outputs,
This may make the attacks more visible from the system operators,
helping them to detect the malicious activities. 
In this sense, the security level can be enhanced by 
using quantization and especially when it is coarse. 
Clearly, there will be a certain tradeoff between
the security level and the control performance since
coarse quantization can in general degrade the system
performance. 

It is important to note that sampled-data control
can introduce vulnerability in the context of zero 
dynamics attacks \cite{ShiBacEun:22}. 
Even if the continuous-time plant
originally is minimum phase without any unstable zero, 
when discretized, 
the socalled sampling zeros will appear and
some of them can be unstable depending on the 
sampling rate \cite{astrom}. Such unstable zeros
can be exploited by the attacker; 
the plant dynamics in the continuous-time domain 
can be excited in such a way that the sampled output
shows no sign of irregular trajectories.

Quantization in networked control has been studied 
in the past two decades from the perspective of reducing
data transmissions for feedback control purposes;
see, e.g., \cite{EliMit:01,IshFra_book:02,Liberzon:03,NaiFagZamEva:07,OkaIsh:17,YouXie:10}.
For systems under DoS attacks, the effects of quantization
have been addressed in \cite{katoDoS,wakaikiDoS,fengDoS}. 
The general implication found there is that using finer 
quantization, which
requires higher data rate for the communication of
control signals, would improve the robustness of the 
system against DoS attacks and vice versa.

To the best of our knowledge, however, 
quantization and their influence on FDI attacks have
not been studied in the literature.
Here, we will provide an analysis on the error in 
the system output caused by quantizing zero dynamics 
attacks as well as their capability to destabilize the
system. Our problem setting is limited to systems under
feedforward control, but 
it will be shown that by taking account of quantization 
effects, attack signals can be generated
resulting in a lower error level in comparison to 
the simple approach of directly quantizing the 
conventional zero dynamics attacks. 
To this end, we propose two quantized attack methods, one
based on dynamic quantizers \cite{dquant,sythdq} and the other
using a modified version of zero dynamics attacks. 
These methods are constructive in that the attacker
can specify the size of tolerable errors in the system
output and generate attack signals accordingly.

This paper is organized as follows: 
Section~\ref{ch:problem} describes the networked control 
system, the input quantizer and the class of the attacks. 
In Section~\ref{ch:dq}, we briefly overview 
the approach of dynamic quantizers, which will be used for
attack signal generation in one of the proposed methods.
In Section~\ref{ch:nonmin}, we explain the two methods 
of quantized attacks for a non-minimum phase 
sampled-data systems and analyze their effects. 
In Section~\ref{ch:example}, we illustrate the effectiveness
of our results via a numerical example.  
In Section~\ref{ch:conclusion}, we provide 
concluding remarks. 
The material of this paper appeared as \cite{KimIsh:22}
in a preliminary form; the current version contains the proofs
of the main results and further discussions. 

\textbf{Notation}: We denote by $\mathbb{R}$, $\mathbb{N}$, 
and $\mathbb{Z}$ the sets of real numbers, natural numbers, and integers, respectively. 
For $d>0$, $d\mathbb{Z}$ is the set of numbers which can be expressed as $dz$ by using an integer $z\in\mathbb{Z}$. 
The space of bounded sequences is denoted by $l_\infty$. 

\section{Problem formulation}\label{ch:problem}
In this section, we formulate the problem of quantized FDI attacks studied in this paper.
\subsection{Sampled-data system under quantized input}\label{se:plant}
\begin{figure}[t]
 \vspace*{1mm}
 \centering
  \includegraphics[width=0.9\linewidth]{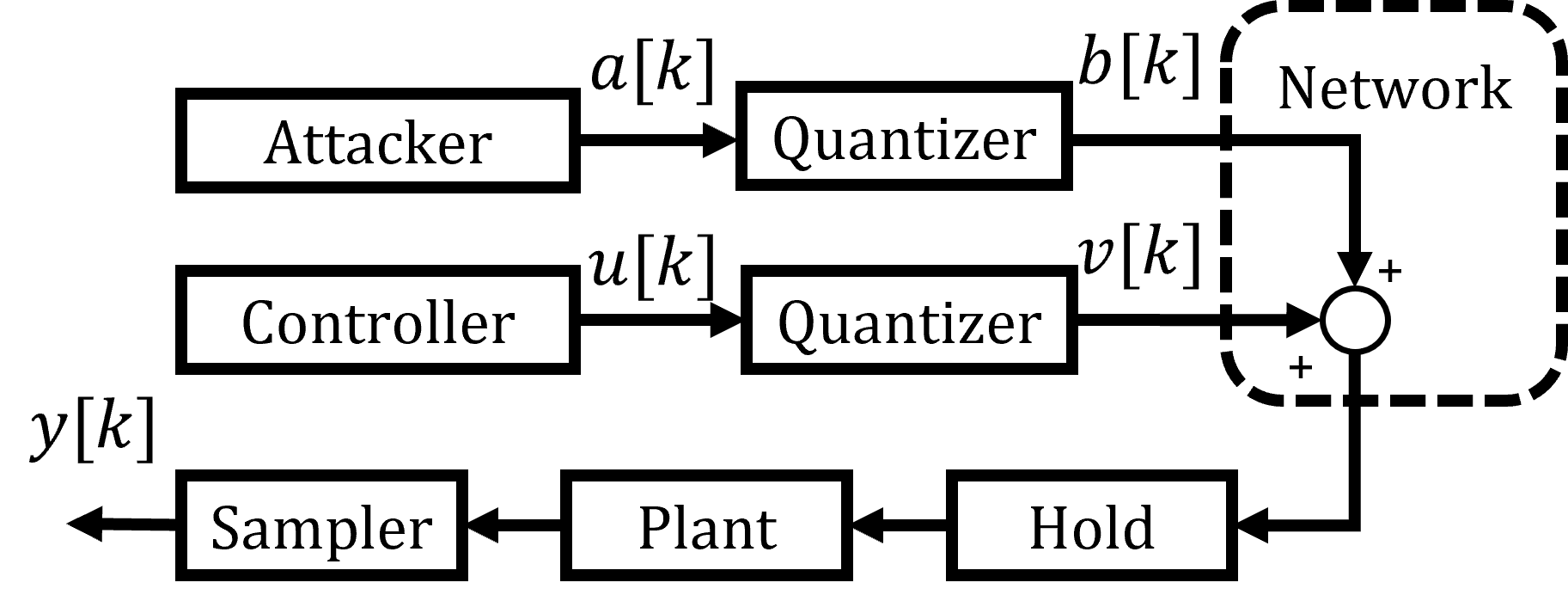}
  \caption{Networked sampled-data control system under attack}
  \label{fig:pr:sys}
\end{figure}

Consider the feedforward sampled-data control system shown in Fig.~\ref{fig:pr:sys}, where the controller is connected with the plant via a network. The control input is generated by the controller and is quantized before it is sent to the plant over the network. A malicious attacker has access to the network and may modify the input signal though the attack signal must use the same quantization scheme as the controller. 

In Fig.~\ref{fig:pr:sys},
the plant is a single-input single-output linear time-invariant (LTI) system whose state-space equation is given by
\begin{equation}
    \begin{split}
        \dot{x}(t)&=A_cx(t)+B_c(v(t)+b(t)),\\
        y(t)&=Cx(t),
    \end{split}\label{eq:pr:csys}
\end{equation}
where $x(t)\in\mathbb{R}^n$ is the state, $y(t)\in\mathbb{R}$ is the output, $A_c\in\mathbb{R}^{n\times n}$, $B_c\in\mathbb{R}^{n\times 1}$, and $C\in\mathbb{R}^{1\times n}$ are the system matrices. Moreover, the inputs $v(t)$ and $b(t)\in d\mathbb{Z}$ are, respectively, the quantized signals of the feedforward input $u(t)$ and the attack signal $a(t)$. The specific quantizer used will be introduced later. We assume that the system is stable and controllable, and 
the initial state is $x(0)=x_0$.

The input $v(t)$ is injected through  the zero-order hold (ZOH) whose sampling period is $T>0$. Denote by $v[k]$ the input calculated by the controller at time $t=kT$. The continuous-time signal applied to the plant is $v(t):=v [k]$ for $t\in[kT,(k+1)T)$. On the other hand, the output $y(t)$ of the plant is sampled at the same rate as the ZOH with sampling period $T$. So the measured output is $y [k] :=y(kT)$. We adopt the notation to write the system state and output of \eqref{eq:pr:csys} driven by input $u(t)$
as $x_u(t)$ and $y_u(t)$, respectively.

As the quantizer, we employ the uniform quantizer $q:\mathbb{R}\rightarrow d\mathbb{Z}$, 
whose width is denoted by $d>0$. 
We use the nearest neighbor quantization towards $-\infty$; 
it maps $\mu\in\mathbb{R}$ to $q(\mu)$ which is the optimal solution obtained as
\begin{equation}
  q(\mu) := \min_{w\in d\mathbb{Z}}\left|w-\mu\right|. 
\label{eqn:qmu}
\end{equation}
In Fig.~\ref{fig:pr:sys}, 
the most simple approach for obtaining the quantized input $v[k]$ is
to directly quantize the input $u[k]$ as 
\begin{equation}   
  v[k] = q(u[k]).
\label{eqn:vk}
\end{equation}
This approach will be referred to as static quantization.

\subsection{Quantized FDI attacks}

As mentioned above, the adversary is capable to modify the quantized control input $v[k]$ to $v[k]+b[k]$ by injecting the quantized attack signal $b[k]$. It is assumed that the adversary has the full information of the plant and the controller including their dynamics whereas the system operator has no information regarding the attacks. The adversary's objective is to disrupt the operation of the system without being detected by the system operator who may be monitoring the sampled system output $y[k]$. 

More specifically, the adversary attempts to generate the quantized 
attack signal $b[k]$ satisfying the following two conditions:\par
(i)~\textit{$\mathbf{\epsilon}$-stealthy condition}: 
The attack signal $\{b[k]\}_{k=0}^{\infty}$ is said to be 
$\epsilon$-stealthy for given $\epsilon >0$ if 
\begin{equation}
    \left|y_{v+b}[k]-y_v[k]\right|\leq\epsilon
\label{eq:pr:estealth}
\end{equation}
for all $k$, where $y_v[k]$ is the original output of the system under the input 
$v[k]$, and $y_{v+b}[k]$ is the output of the attacked system.\par
(ii)~\textit{$\{H_k\}_{k=0}^{\infty}$-disruptive condition}: Given a sequence of nonnegative numbers $\{H_k\}_{k=0}^{\infty}$, the attack signal $\{b[k]\}_{k=0}^{\infty}$ is said to be  $\{H_k\}_{k=0}^{\infty}$-disruptive if there exists a time 
sequence $\{t_k\}_{k=0}^{\infty}$ such that for any $k$, it holds
\begin{equation}
    \left\|x_{v+b}(t_k)-x_v(t_k)\right\|_2\geq H_k.
\label{eq:pr:disrupt}
\end{equation}

The problem of quantized zero dynamics attacks addressed in the paper can be stated as follows:

\textit{Problem}: Consider the quantized networked system in Fig.~\ref{fig:pr:sys}. 
Let the positive scalar $\epsilon$ and the sequence $\{H_k\}_{k=0}^{\infty}$ of nonnegative numbers be given. Then find the quantized attack signal $\{b[k] \}_{k=0}^{\infty}$ which satisfies both the $\epsilon$-stealthy and the $\{H_k\}_{k=0}^{\infty}$-disruptive conditions.

An important aspect of our problem setting is that we examine the case where 
the sampled-data system has unstable zeros in the discrete-time domain after discretization.
As we will see, this may hold even if the original continuous-time plant  
does not have unstable zeros \cite{ShiBacEun:22}.

If the attacker can inject real-valued signals (i.e., without quantization), attack methods satisfying the two conditions above have been proposed, which are known as the zero dynamics attacks  \cite{TeiShaSan:15,ParShiLee:19}. This type of attacks takes advantage of system zeros. Notably, if the system has unstable zeros, states may diverge and it is difficult to detect such attacks from the output since it will remain almost the same with or without the attack.\par
However, if we place an input quantizer, the system state/output under attacks should be influenced. Consequently, the stealthy and disruptive properties may be lost to some extent. In fact, if the attack signals are statically quantized, then the output may receive direct and large influences as we will see later. Hence, quantization may be used as a means to detect attacks on the system. Note that however that quantizing the input would reduce the control performance at the same time. So there is a tradeoff between the attack detection and control performance. The objective of this paper is to demonstrate that different methods for quantization of attack signals can result in different properties and in particular there are ways to reduce the effects caused by quantizers.

\subsection{Zero dynamics attacks}\label{se:zda}

The sampled-data system \eqref{eq:pr:csys} can be expressed as the following discrete-time linear system under ZOH and sampling:
\begin{equation}
    \begin{split}
        x[k+1]&=Ax[k]+B(v[k]+b[k]),\\
        y[k]&=Cx[k],
    \end{split}\label{eq:pr:dsys}
\end{equation}
where $x[k]:=x(kT)$ is the sampled state, $A:=\text{e}^{A_cT}$, and $B:=\int_0^T\text{e}^{A_c\tau}d\tau B_c$. We assume that this system is controllable and the vector product $CB$ is nonzero, which means the relative degree of this discretized system is $1$; these properties hold for almost any $T>0$ \cite{astrom}. \par
Note that when a continuous-time system is discretized, zeros which do not exist in the original system may appear. These are called the sampling zeros. As mentioned above, regardless of the relative degree of the original system \eqref{eq:pr:csys}, the discretized system \eqref{eq:pr:dsys} has relative degree of $1$ for almost any sampling period $T$. It is known that when $T$ is small, some of them will be unstable \cite{astrom}. Hence, even if the original system has no unstable zero, there is a chance that discretization will introduce some. In such a case, if the input is not quantized, then the zero dynamics attacks proposed in \cite{TeiShaSan:15,ParShiLee:19} can be applied. We note that such attacks can have arbitrarily small effects on the output by taking the initial state accordingly.

In view of the above, we impose the following assumption throughout the paper
unless otherwise noted:

\begin{assump}\label{assump:1}
The discretized plant \eqref{eq:pr:dsys} has at least one nonminimum phase zero 
(i.e., unstable zero).
\end{assump}

We summarize how to construct zero dynamics attack signals under this assumption as follows: 
Here, we consider the case where the control input and the attack signal are not quantized, i.e., $v[k]=u[k]$ and $b[k]=a[k]$. According to \cite{TeiShaSan:15}, we can construct the attack signal $\{a[k]\}_{k=0}^{\infty}$ by
\begin{equation}
    \begin{split}
        z[k+1]&=\left(A-\frac{BCA}{CB}\right)z[k],\\
        a[k]&=-\frac{CA}{CB}z[k],
    \end{split}\label{eq:pr:zda}
\end{equation}
where $z[k]\in\mathbb{R}^n$ is the state whose initial value satisfies 
$z[0]\neq 0$ and $z[0]\in\ker(C)$ with small 
$\left\|z[0]\right\|_2$.
Note that the nonminimum phase property of \eqref{eq:pr:dsys} implies that the system matrix $A-{BCA}/{(CB)}$ is unstable. The core idea of zero dynamics attacks is that this attack signal $a[k]$ will make the system state diverge but keep it in the undetectable region. Regarding this type of attacks, the following result is fundamental \cite{TeiShaSan:15}.
\begin{lem}\label{lem:zda}
Consider the system in \eqref{eq:pr:dsys}. For any $\epsilon>0$, there exist attack signals such that they are $\epsilon$-stealthy and $\left\|x_{u+a}[k]-x_u[k]\right\|_2$ diverges as $k\rightarrow\infty$.
\end{lem}

\section{Dynamic quantization}\label{ch:dq}
In this section, we briefly introduce the method called dynamic quantization from the works \cite{dquant,dqseri,sythdq}.
This will later be applied for the quantization of attack signals.

\begin{figure}[t]
    \begin{minipage}[t]{1\linewidth}
        \centering        
\includegraphics[width=0.7\linewidth]{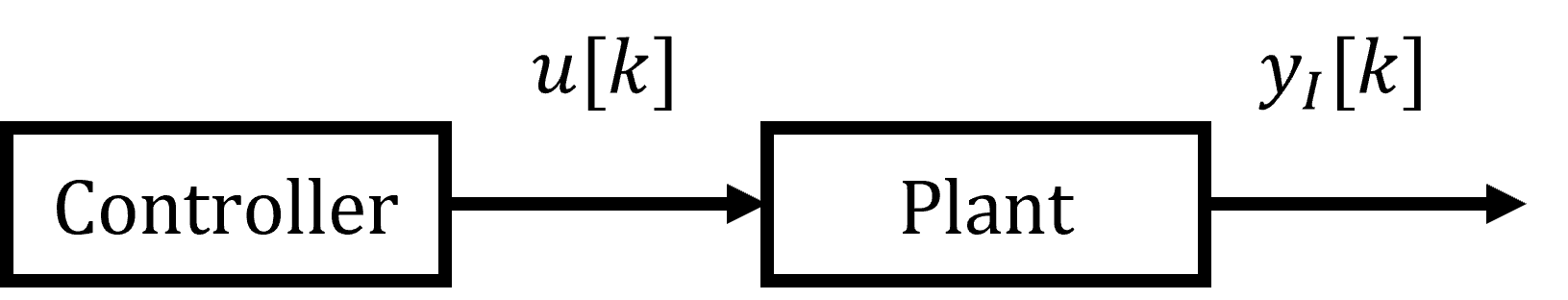}
\vspace*{1mm}
        \caption{Controller and plant}
        \label{fig:dq:osys}
    \end{minipage}
    \vspace{0.04\linewidth}\\
    \begin{minipage}[t]{1\linewidth}
        \centering
     \includegraphics[width=1\linewidth]{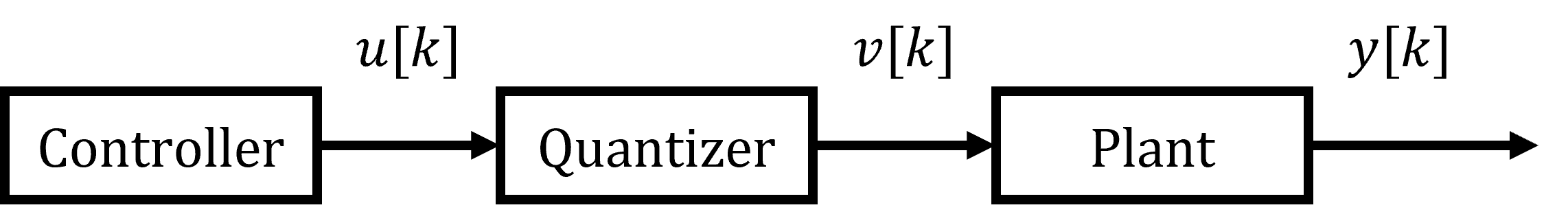}
\vspace*{-2mm}
        \caption{Controller and plant with an input quantizer}
        \label{fig:dq:qsys}
    \end{minipage}
\end{figure}

We consider the two feedforward discrete-time systems shown in Figs.~\ref{fig:dq:osys} and \ref{fig:dq:qsys}, 
where the controllers apply the same input $u[k]$ 
to both systems. The plant is from \eqref{eq:pr:dsys} 
without the attack signal, i.e., $b[k]\equiv0$. 
The difference between the two systems is that the 
first system in Fig.~\ref{fig:dq:osys} is the original 
one with the ideal output denoted by $y_I[k]$ whereas 
in the second system in Fig.~\ref{fig:dq:qsys}, 
the input is quantized, resulting in the output 
$y[k]$ with some deviations from $y_I[k]$. 
The problem is to find a quantizer with dynamics 
to minimize the error $y[k]-y_I[k]$ based on the past 
inputs $u[k]$ and their quantized values $v[k]$.
In particular, the measure of output error is defined as
\begin{equation}
    E_q := \sup_{u,k}\left|y[k]-y_I[k]\right|.
\label{eq:dq:error}
\end{equation}\par

Before we proceed to dynamic quantizers, we can
look at the case of static quantization 
when the input $u[k]$ is directly quantized as $v[k]=q(u[k])$
as in \eqref{eqn:vk}.
In this case, it is known that 
the output error \eqref{eq:dq:error} can be obtained as 
\begin{equation}
 E_q
  = \frac{1}{2}\sum\limits_{k=0}^\infty\left|CA^kB\right| d.
\label{eq:dq:staticeps}
\end{equation}
Since the plant is assumed to be stable, this value is
finite. This formula suggests that for systems with
slower convergence, quantization has a larger impact
on the output, and moreover this error is linear with
respect to the quantization width $d$.

\subsection{Basic structure of dynamic quantizers}\label{se:dq}
In this subsection, we assume that the plant \eqref{eq:pr:dsys} is minimum phase (i.e., 
Assumption~\ref{assump:1} is not imposed).
The dynamic quantizer is given by a linear system whose output is quantized by 
the uniform quantizer \eqref{eqn:qmu}:
\begin{equation}
    \begin{aligned}
    \xi[k+1]&=E\xi[k]+F_1v[k]+F_2u[k],\\
    v[k]&=q(G\xi[k]+Hu[k]),
    \end{aligned}
    \label{eq:dq:dquant}
\end{equation}
where $\xi[k]\in\mathbb{R}^{n_q}$ is the state and its initial value is taken as $\xi[0]=0$;
the system matrices are of appropriate sizes. 
We would like to design the dynamic quantizer 
so as to minimize the output error in \eqref{eq:dq:error}.
Note that the static quantization in \eqref{eqn:vk} 
is a special case, corresponding to $G=O$ and $H=I$.

The dynamic quantizer is known to be optimal if
\begin{equation}
  \begin{split}
     E&=A,~~F_1=-F_2=B,~~
     G=-\frac{CA}{CB},~~H=I,
    \end{split}
\label{eq:dq:param}
\end{equation}
and then the minimum output error is given by \cite{dquant}
\begin{equation}
    E_q = \frac{1}{2}\left|CB\right|d.
\label{eq:dq:erroropt}
\end{equation}

We observe that with dynamic quantization,
the output error may become smaller than
that for the static case in \eqref{eq:dq:staticeps}.
It is noted that $E_q$ in \eqref{eq:dq:erroropt} is
independent of the system matrix $A$. So different from
the static case, the output error does not depend on
the convergence rate of the plant.

However, we must note that this design method can be 
applied only to minimum phase systems. 
In fact, when the system \eqref{eq:pr:dsys} has 
unstable zeros, the dynamic quantizer outlined 
above becomes unstable. The case for nonminimum 
phase systems is discussed next. 

\subsection{Dynamic quantizer based on serial decomposition}\label{se:dqseri}
We describe the dynamic quantization approach of 
\cite{dqseri} based on decomposition of the plant. 
Suppose that the system \eqref{eq:pr:dsys} has unstable zeros. We denote by $P$ the system from the quantized input $v[k]$ to the output $y[k]$ in \eqref{eq:pr:dsys}. This is decomposed serially to subsystems $P_s$ and $P_u$ as $P=P_s\cdot P_u$ such that the following hold:
\begin{enumerate}
    \item $P_s$ is stable, minimum phase and strictly proper, and
    \item $C_sB_s\neq0$,
\end{enumerate}
where the realizations of $P_s$ and $P_u$ are, respectively, given by $(A_s,B_s,C_s,0)$ and $(A_u,B_u,C_u,D_u)$. Note that the minimum phase property of $P_s$ implies that $A_s-{B_sC_sA_s}/{(C_sB_s)}$ is a stable matrix.

Dynamic quantization based on serial decomposition generates the quantized signal $v[k]$ as follows:
\begin{equation}
    \begin{aligned}
    \xi[k+1]
      &= A_s\xi[k]+B_sv[k]-B_su[k], \\
    v[k]
      &= q\left(-\frac{C_sA_s}{C_sB_s}\xi[k]+u[k]\right),
    \end{aligned}
\label{eq:dq:dqseri}
\end{equation}
where $\xi[0]=0$.
Notice that this is the optimal quantizer for the system $P_s$
from \eqref{eq:dq:param} and \eqref{eq:dq:dquant}.
The output error in \eqref{eq:dq:error} for this case 
can be expressed as
\begin{equation}
E_q
 = \frac{1}{2}\left\|P_u\right\|_{i_\infty}
    \left|C_sB_s\right|d,
\label{eq:dq:dqserierroropt}
\end{equation}
where $\left\|P_u\right\|_{i\infty}$ is the induced-$l_\infty$ norm of the subsystem $P_u$ given by
    $\left\|P_u\right\|_{i\infty}
      :=\sup_{r\in l_\infty,r\neq0}
          {\left\|P_ur\right\|_\infty}/{\left\|r\right\|_\infty}$.
This is the contribution of $P_u$ to the output error 
in \eqref{eq:dq:dqserierroropt}.

Note that in this method, there is some freedom in the choices of the subsystems $P_s$ and $P_u$. One difficulty is that if $P$ has multiple unstable zeros, then the optimal way of decomposition is not known in general. One exception is when the plant $P$ has only one unstable zero. In that case, the optimal decomposition and output error can be obtained explicitly. Let $\lambda$ be the unstable zero $(|\lambda|>1)$ of $P$. Then, $P_u$ should be taken in the transfer function form as $P_u(z)=(z-\lambda)/z$. The optimal output error 
in \eqref{eq:dq:dqserierroropt} can be obtained as
\begin{align}
    E_q &= \frac{1}{2}(1+\left|\lambda\right|)\left|C_sB_s\right|d 
        = \frac{1}{2}(1+\left|\lambda\right|)\left|C B\right|d.
\label{eq:dq:dqseriopt}
\end{align}
The second equality can be shown by the serial decomposition 
and by the choice of $P_u$, resulting in the direct feedthrough term 
$D_u=1$ in its state-space form.

\section{Quantized attacks for non-minimum phase systems}\label{ch:nonmin}
In this section, we discuss quantized attacks for the case when the sampled-data system \eqref{eq:pr:csys} has unstable zeros. Concretely, we consider three types of methods as follows:

(i)~The first method is simple, where zero dynamics attacks as described in Section~II-C are generated and quantized directly using the static quantizer.\par
(ii)~The second approach employs the dynamic quantizer outlined above to modify the zero dynamics attacks in real values. We can show that this method can achieve smaller output error than statically quantizing the zero dynamics attacks as in the first method. However, it is affected by the number of unstable zeros and their values. \par
(iii)~The third method quantizes attack signals which are slightly different from the conventional zero dynamics attacks and may induce small but bounded influence on the system output. This signal can be constructed without the influence of unstable zeros. Moreover, as we will see in the analysis as well as numerical examples, this method is capable to outperform the other two methods in terms of the output error.\par
In this section, we look at these three attack methods and analyze the resulting output errors. 
\subsection{Quantized attack method 1 via static quantization}\label{se:sqzda}
We start with the simplest quantization approach where the attack signal $a[k]$ generated by \eqref{eq:pr:zda} is directly 
quantized in a static manner. That is, the attack
signal $b[k]$ to be injected to the quantized control input $v[k]$ 
in Fig.~1 is given by
\begin{equation}
    b[k] = q(a[k]).
    \label{eq:no:bqa}
\end{equation} 
For this method, we obtain the following result.
\begin{prop}
For the quantized networked system under attacks in Fig.~\ref{fig:pr:sys}, suppose that the quantization width $d$ is given and the scalar $\epsilon$ satisfies
\begin{equation}
    \epsilon
     > \frac{1}{2}\sum\limits_{k=0}^\infty\left|CA^kB\right|d.
\label{eq:no:staticeps}
\end{equation}
Moreover, suppose that the attack signal generated by \eqref{eq:pr:zda} is statically quantized as in \eqref{eq:no:bqa}. Then, for any sequence $\{H_k\}_{k=0}^{\infty}$, there exists a quantized attack signal $\{b[k]\}_{k=0}^{\infty}$ that is $\epsilon$-stealthy and $\{H_k\}_{k=0}^{\infty}$-disruptive.
\end{prop}

\begin{proof}
This can be shown easily, and we provide only the outline
of the proof. First, we upper bound the output error as
\begin{multline*}
   \left|y_{v+b}[k]-y_v[k]\right|\\
     \leq \left|y_{v+b}[k]-y_{v+a}[k]\right|
           + \left|y_{v+a}[k]-y_v[k]\right|.
\end{multline*}
The first term on the right-hand side is less than or equal to $({1}/{2})\sum_{k=0}^\infty\left|CA^kB\right|d$ from \eqref{eq:dq:staticeps}. By Lemma.~1 and \eqref{eq:no:staticeps}, we can choose the attack signal $\{a[k]\}_{k=0}^{\infty}$ which is $\left(\epsilon-({1}/{2})\sum_{k=0}^\infty\left|CA^kB\right|d\right)$-stealthy. Then, the right-hand side of the above inequality is less than or equal to $\epsilon$. Therefore the attack $\{b[k]\}_{k=0}^{\infty}$ is $\epsilon$-stealthy.\par
Next, we look at the difference in the states. We have
\begin{align*}
  &\left\| x_{v+b}[k]-x_v[k]\right\|_2\\
   &~~\geq \left\| x_{v+a}[k]-x_v[k]\right\|_2
       - \left\| x_{v+b}[k]-x_{v+a}[k]\right\|_2.
\end{align*}
On the right-hand side, the first term diverges as $k\rightarrow\infty$. The second term on the right-hand side coincides with the state which is driven by the quantization error; since the quantization error is bounded and the system \eqref{eq:pr:dsys} is stable, this term is also bounded. It means that $\left\| x_{v+b}[k]-x_v[k]\right\|_2$ diverges as $k\rightarrow\infty$. Therefore, we conclude that the attack $\{b[k]\}_{k=0}^{\infty}$ is $\{H_k\}_{k=0}^{\infty}$-disruptive.
\end{proof}

\subsection{Quantized attack method 2 via dynamic quantization}\label{se:dqzda}

In this subsection, we propose a method which modifies the continuous zero dynamics attacks in Section \ref{ch:problem} by using dynamic quantization from the previous section. This method adopts dynamic quantizer based on serial decomposition in Section \ref{se:dqseri}. \par

Specifically, we apply \eqref{eq:dq:dqseri} to generate attack signals after obtaining the serial decompositions $P_s$ and $P_u$ of the system \eqref{eq:pr:dsys} satisfying conditions outlined there. The effects to output signals caused by this attack is determined by the choices of 
$P_s$ and $P_u$. As we discussed earlier, while this method can be utilized for systems with any number of unstable zeros, the optimal value for the output error is known only for the case of one unstable zero. The following theorem is the first main result of this paper stating that dynamic quantizer can be efficient in reducing the error due to quantized attacks. 
\begin{theo}\label{theo:dqzda}
For the quantized networked system under attacks in Fig.~\ref{fig:pr:sys}, suppose that the plant is decomposed serially as described in Section \ref{se:dqseri} and $\epsilon$ satisfies
\begin{equation}
    \epsilon
     > \frac{1}{2}\left\|P_u\right\|_{i\infty}\left|C_sB_s\right|d,
\label{eq:no:eps}
\end{equation}
where $B_s$ and $C_s$ are the system matrices of $P_s$. Then, for any sequence $\{H_k\}_{k=0}^\infty$, there exists a quantized attack signal $\{b[k]\}_{k=0}^\infty$ generated by \eqref{eq:pr:zda} and dynamic quantization that is $\epsilon$-stealthy and $\{H_k\}_{k=0}^\infty$-disruptive.
\end{theo}\par

\begin{proof}
By Lemma \ref{lem:zda}, we can construct the attack signal $\{a[k]\}_{k=0}^{\infty}$ such that it is $\left(\epsilon-({1}/{2})\left\|P_u\right\|_{i\infty}\left|C_sB_s\right|d\right)$-stealthy and $\left\|x_{u+a}[k]-x_u[k]\right\|_2$ diverges. Consider the attack $\{b[k]\}_{k=0}^{\infty}$ which is made from $\{a[k]\}_{k=0}^{\infty}$ and the dynamic quantizer in Section \ref{se:dqseri}. We verify that this attack satisfies both $\epsilon$-stealthy and $\{H_k\}_{k=0}^{\infty}$-disruptive properties.\par
We can upper bound the output error as
\begin{align}
  & |y_{v+b}[k]-y_v[k]| \notag\\
  &~~~\leq|y_{v+b}[k]-y_{v+a}[k]|
          +| y_{v+a}[k]-y_v[k]|\label{eq:no:ineqy}
\end{align}
for every $k$. Since the system is linear, the first term on the right-hand side coincides with $|y_b[k]-y_a[k]|$. By the bound on the output error \eqref{eq:dq:dqserierroropt} for the dynamic quantizer, this term can be bounded as
\begin{equation}
    \sup\limits_k|y_b[k]-y_a[k]|
     \leq \frac{1}{2}\left\|P_u\right\|_{i\infty}|C_sB_s|d.\notag
\end{equation}
By the construction of $\{a[k]\}_{k=0}^{\infty}$, the second term on the right-hand side of \eqref{eq:no:ineqy} satisfies the following inequality:
\begin{equation}
    \sup\limits_k|y_{v+a}[k]-y_v[k]|
       \leq \epsilon - \frac{1}{2}\left\|P_u\right\|_{i\infty}|C_sB_s|d.\notag
\end{equation}
Hence, from \eqref{eq:no:ineqy}, we have $\sup_k| y_{v+b}[k]-y_v[k]|\leq\epsilon$. Consequently, the attack $\{b[k]\}_{k=0}^\infty$ is $\epsilon$-stealthy.\par
Next, we look at the difference in states. We have
\begin{multline}
    \left\| x_{v+b}[k]-x_v[k]\right\|_2\\\geq\left\| x_{v+a}[k]-x_v[k]\right\|_2-\left\| x_{v+b}[k]-x_{v+a}[k]\right\|_2.\label{eq:no:statediff}
\end{multline}
Because of the choice of $\{a[k]\}_{k=0}^{\infty}$, the first term on the right-hand side diverges as $k\rightarrow\infty$. Since the system is linear, it follows that $x_{v+b}[k]-x_{v+a}[k]=x_b[k]-x_a[k]$. The state-space equation of the plant and the dynamic quantizer can be written as
\begin{equation}
\begin{bmatrix}
x_{b}[k+1] \\
\xi[k+1]\\
x_{a}[k+1]
\end{bmatrix}\hskip-3pt=\hskip-3pt\left[\begin{array}{@{\hskip-1pt}c@{\hskip1pt}c@{\hskip1pt}c@{\hskip-1pt}}
A&O&O\\
O&A_s&O\\
O&O&A 
\end{array}\right]\hskip-5pt\begin{bmatrix}
x_{b}[k] \\
\xi[k]\\
x_{a}[k]
\end{bmatrix}\hskip-2pt+\hskip-2pt\begin{bmatrix}
Bb[k] \\
B_s(b[k]-a[k])\\
Ba[k]
\end{bmatrix}.\label{eq:no:xbxixa}
\end{equation}
Let $w_q[k]$ be
\begin{equation*}
 w_q[k]
  := q\left(-\frac{C_sA_s}{C_sB_s}\xi[k]+a[k]\right)
      - \left(-\frac{C_sA_s}{C_sB_s}\xi[k]+a[k]\right).
\end{equation*}
Then, the quantized attack is expressed as $b[k]=a[k]+w_q[k]-{C_sA_s}/({C_sB_s})\xi[k]$. 
Note that $\left\|w_q[k]\right\|_\infty\leq d/2$. Also, let 
\begin{equation*}
    \bar{A}:=\begin{bmatrix}
    A & -\frac{BC_sA_s}{C_sB_s} \\
    O & A_s-\frac{B_sC_sA_s}{C_sB_s}
    \end{bmatrix}.
\end{equation*}
Since the system \eqref{eq:pr:dsys} is stable and the subsystem $P_s$ is minimum phase, 
the matrix $\bar{A}$ is also stable. 

The state-space equation in \eqref{eq:no:xbxixa} can be rewritten as
\begin{equation}
\begin{bmatrix}
x_{b}[k+1] \\
\xi[k+1]\\
x_{a}[k+1]
\end{bmatrix}\hskip-3pt=\hskip-3pt\left[\begin{array}{@{\hskip-1pt}c|c@{\hskip-1pt}}
\raisebox{-0.5em}{\mbox{\LARGE $\bar{A}$}} & \begin{matrix}
O \\ O
\end{matrix}
\\
\hline
 \begin{matrix}
O & O
\end{matrix} & A
\end{array}\right]\hskip-5pt\begin{bmatrix}
x_{b}[k] \\
\xi[k]\\
x_{a}[k]
\end{bmatrix}+\begin{bmatrix}
B(a[k]+w_q[k]) \\
B_sw_q[k]\\
Ba[k]
\end{bmatrix}.\notag
\end{equation}
From this system, we can obtain the following 
one of reduced dimension that takes
$x_{b}[k]-x_a[k]$ as the state:
\begin{equation}
\begin{bmatrix}
 x_{b}[k+1]-x_a[k+1]\\
 \xi[k+1]
\end{bmatrix}
 = \bar{A} 
    \begin{bmatrix}
       x_{b}[k]-x_a[k]\\
       \xi[k]
    \end{bmatrix}
  +  \begin{bmatrix}
        B\\
        B_s
     \end{bmatrix} w_q[k].
\end{equation}
As $\bar{A}$ is stable, 
the sequence $x_{b}[k]-x_a[k]$ is bounded, implying that the second 
term on the right-hand side of \eqref{eq:no:statediff} is bounded. 
Consequently, $\left\| x_{v+b}[k]-x_v[k]\right\|_2\rightarrow\infty$ as $k\rightarrow\infty$. Therefore, we conclude that the quantized attack signal $\{b[k]\}_{k=0}^\infty$ is $\{H_k\}_{k=0}^{\infty}$-disruptive.
\end{proof}

\subsection{Quantized attack method 3 via $\epsilon$-stealthy approach}\label{se:calc}

The third method employs a continuous attack signal 
which is a slightly modified version of the zero 
dynamics attack in \eqref{eq:pr:zda} and then quantizes 
it directly. 

To this end, for given $\epsilon>0$, let us consider 
the attack signal $\{a[k]\}_{k=0}^{\infty}$ generated by
\begin{equation}
    \begin{split}
        z[k+1]&=\left(A-\frac{BCA}{CB}\right)z[k]+\frac{B}{CB}\epsilon,\\
        a[k]&=-\frac{CA}{CB}z[k]+\frac{\epsilon}{CB},
    \end{split}\label{eq:no:zdag}
\end{equation}
where $z[0]=0$. We can easily show that this system 
is unstable and the state difference difference $x_{u+a}[k]-x_u[k]$ is equal to $z[k]$. 
It means that $y_{u+a}[k]-y_u[k]=Cz[k]$ and this value 
coincides with $\epsilon$ for $k\geq1$. 
Hence, this attack is $\epsilon$-stealthy in the sense of \eqref{eq:pr:estealth} (without quantization).

The system \eqref{eq:no:zdag} motivates us to generate attack
signals using a system which has constant input proportional 
to $\epsilon$ and then to quantize the signal. 
This can be achieved by introducing the following quantized system:
\begin{equation}
 \begin{split}
   z[k+1]&=Az[k]+Bb[k],\\
        b[k]&=q\left(-\frac{CA}{CB}z[k]+\frac{\sign(CB)\epsilon}{CB}-\frac{d}{2}+\Delta\right),
    \end{split}\label{eq:no:dzdag}
\end{equation}
where $z[0]=0$ and 
$\Delta$ is a sufficiently small positive number. \par
Now, we are ready to state our second main result of the paper. 
\begin{theo}\label{theo:2}
For the quantized networked system under attacks in Fig.~\ref{fig:pr:sys}, suppose that $\epsilon$ is taken as $\epsilon \geq |CB|d$. Then, for any sequence $\{H_k\}_{k=0}^{\infty}$, the quantized attack $\{b[k]\}_{k=0}^{\infty}$ generated by \eqref{eq:no:dzdag} is $\left(\epsilon + |CB|\Delta\right)$-stealthy and $\{H_k\}_{k=0}^{\infty}$-disruptive. 
\end{theo}
\begin{proof}
It is straightforward to show that by \eqref{eq:no:zdag}, the differences 
in the states and outputs can be written as $x_{v+b}[k]-x_v[k]=z[k]$ 
and $y_{v+b}[k]-y_v[k]=Cz[k]$, 
respectively. Let  
\[
  w[k]
   := \left(-\frac{CA}{CB}z[k]+\frac{\sign(CB)\epsilon}{CB}+\Delta\right)
       - b[k]. 
\]
Then, \eqref{eq:no:dzdag} becomes
\begin{multline}
    z[k+1]=\left(A-\frac{BCA}{CB}\right)z[k]\\+B\left(\frac{\sign(CB)\epsilon}{CB}+\Delta-w[k]\right),
\end{multline}
where $0<w[k]\leq d$. Since $z[0]=0$, we have 
\begin{align}
&z[k]\notag\\
&~~= \sum_{l=0}^{k-1}\left(A-\frac{BCA}{CB}\right)^{k-l-1}
   B\left(\frac{\sign(CB)\epsilon}{CB}+\Delta-w[l]\right).
\label{eq:no:zk}
\end{align}
Note that it holds $C\left(A-{BCA}/({CB})\right)=0$, and thus, $Cz[k]=\sign(CB)\epsilon+CB\Delta-CBw[k-1]$ for all $k\geq1$. 
Since $\epsilon\geq|CB|d$, we have
\[
  -\left(\epsilon+|CB|\Delta\right)
    \leq|Cz[k]|\leq\epsilon+|CB|\Delta. 
\]
It means that the attack $\{b[k]\}_{k=0}^{\infty}$ is $\left(\epsilon+|CB|\Delta\right)$-stealthy.\par
Next, to establish that $\{b[k]\}_{k=0}^{\infty}$ is $\{H_k\}_{k=0}^{\infty}$-disruptive, we must show that $z[k]$ diverges, which implies that $\left\|x_{v+b}[k]-x_v[k]\right\|_2$ also diverges. Let 
\[
  w'[k]:=\frac{\sign(CB)\epsilon}{CB}+\Delta-w[k].
\] 
Now, based on \eqref{eq:no:zk}, we can rewrite the system of $z[k]$ and
obtain
\begin{equation*}
   z[k+1] = \left(A-\frac{BCA}{CB}\right) z[k] + B w'[k].
\end{equation*}
Since $z[0]=0$, we must make sure that in this system, the input $w'[k]$ enters and excites the unstable mode. \par 
Recall that $\left(A,B\right)$ is controllable, and thus $\left(A-{BCA}/{(CB)}, B\right)$ is also controllable. Hence, it follows that the input $w'[k]$ enters every mode of the system. Finally, we can confirm that $w'[k]$ is not zero at all times. By assumption, $\epsilon\geq|CB|d$, and this implies that $w'[k]\geq\Delta$ at each $k$. Therefore, the state $z[k]$ diverges.
\end{proof}

We remark that to achieve the disruptive property, the assumption that $\epsilon \geq|CB|d$ is critical for the quantized attack \eqref{eq:no:dzdag}. This is because the initial state is $z[0]=0$, if $\epsilon<|CB|d$, then $z[k]\equiv0$. This will further result in $b[k]\equiv0$, that is, no quantized attack.
Furthermore, compared with the attack approach based on dynamic quantization of
Section \ref{se:dqzda}, the attack \eqref{eq:no:dzdag} of this method has 
an advantage in the specific case when the plant has only one unstable zero:
Then, the output error of this approach 
can be made strictly smaller than that for dynamic quantization.
 whose $\epsilon$ 
is lower bounded as shown in \eqref{eq:no:eps}. 

This holds because with sufficiently small $\Delta$, we have
\begin{equation*}
  \frac{1}{2}\left|CB\right|
    \left(1+|\lambda|\right)d
   >\left|CB\right|\left(d+\Delta\right),
\end{equation*}
where $\lambda$ is the unstable zero with $|\lambda|>1$.
We will confirm this difference between the two methods
in numerical examples presented in the next section.

\section{Numerical Example}\label{ch:example}
In this section, we illustrate the results of our paper through numerical simulations. \par
Consider the following stable continuous-time system with quantized input under attack \cite{enhancement}:
\begin{equation*}
    \begin{split}
        \dot{x}(t)&=\begin{bmatrix}
        0 & 0 & 1 & 0\\
        0 & 0 & 0 & 1 \\
        -2 & 1 & -1 & 1\\
        1 & -1 & 1 & -1
        \end{bmatrix}x(t)+\begin{bmatrix}
        0\\0\\0\\1
        \end{bmatrix}(v(t)+b(t)),\\
        y(t)&=\begin{bmatrix}
        1&0&0&0
        \end{bmatrix}x(t).
    \end{split}
\end{equation*}
This system has one stable zero whose value is $-1$. When discretized using sampling period $T=0.5$, the system has three zeros at $-3.21$, $-0.24$, and $0.61$ in the discrete-time domain. In particular, there is one unstable zero $\lambda=-3.21$. We set the initial state as $x_0=0$ and the input as $u[k]=\sin0.05\pi k+0.5\cos0.025\pi k$. Let the quantization width be  $d=1$. We compare the three methods for quantized attacks from Section \ref{ch:nonmin}:~(i)~Static quantization, (ii)~dynamic quantization, and (iii)~$\epsilon$-stealthy approach.

\begin{figure}[t]
    \begin{minipage}[h]{1\linewidth}
        \centering
        \includegraphics[width=1\linewidth]{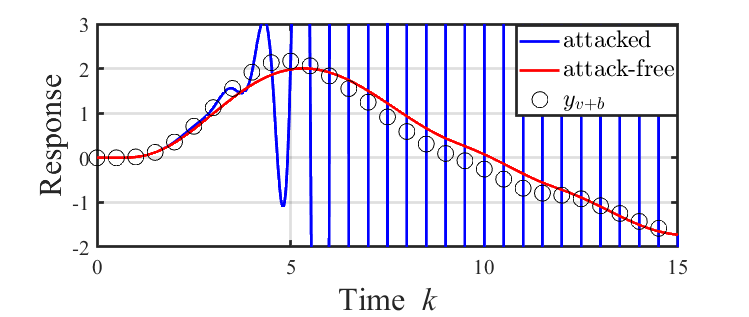}
        \subcaption{Outputs of the system in continuous time and discrete time  \label{fig:no:respnonminsta}}
    \end{minipage}\\
    \begin{minipage}[h]{1\linewidth}
    \vspace*{4mm}
        \centering
        \includegraphics[width=1\linewidth]{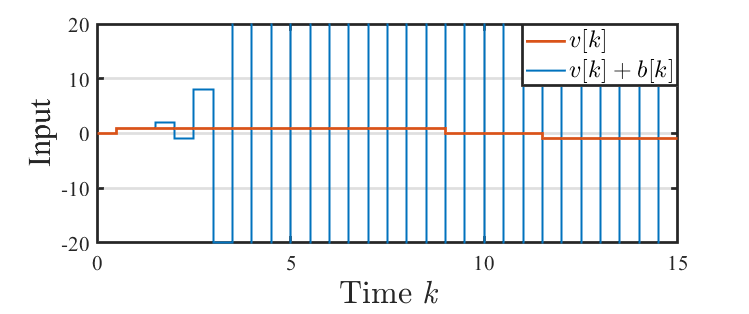}
        \subcaption{Quantized control inputs before and after attacks  \label{fig:no:stazda}} 
    \vspace*{1mm}
    \end{minipage}
    \caption{Time responses of the system under the statically quantized zero dynamics attack. \label{fig:no:stzdaresult}}
\end{figure} 

(i) First, a static quantizer is used for zero dynamics attacks as described 
in Section~\ref{se:sqzda}. 
In Fig.~\ref{fig:no:respnonminsta}, the time responses of the output of the plant 
are shown. The blue line is the continuous-time output (before sampling) under attack while 
the black circles indicate the output at sampling times $t=kT$ for $k\in\mathbb{Z}_+$. 
The red line is the continuous-time output when no attack is present. 
Fig.~\ref{fig:no:stazda} displays the quantized signals of the control input and the sum of the input and the attack. We observe that the input and thus the plant states are diverging, but this is difficult to detect from the sampled output, which is fairly close to the expected behavior of the output under normal conditions.  From \eqref{eq:no:staticeps}, the theoretical bound on the error between the two output signals is $\sum_{k=0}^\infty\left|CA^kB\right|{d}/{2}=5.717$, which is clearly satisfied in the simulation results.

\begin{figure}[t]
    \begin{minipage}[h]{1\linewidth}
        \centering
        \includegraphics[width=1\linewidth]{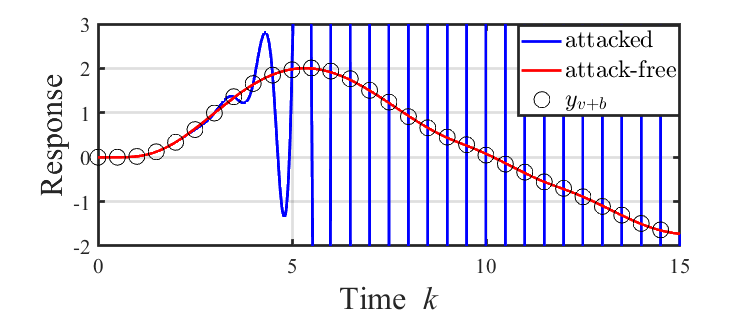}
        \subcaption{System responses   \label{fig:no:respnonmin1}}
    \end{minipage}\\
\vspace*{1mm}
    \begin{minipage}[h]{1\linewidth}
       \vspace*{4mm}
        \centering
        \includegraphics[width=1\linewidth]{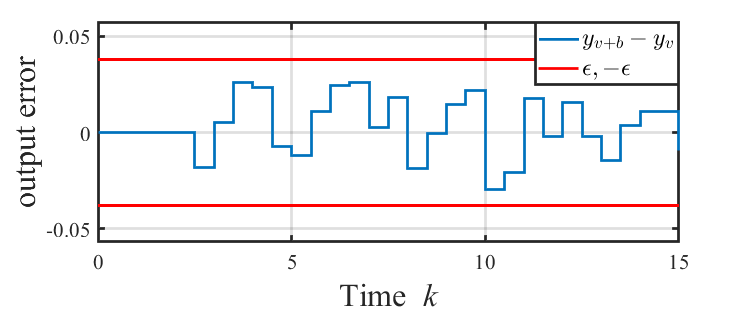}
        \subcaption{Difference in the outputs of the system    \label{fig:no:diffy1}}
    \end{minipage}\\
\vspace*{1mm}
    \begin{minipage}[h]{1\linewidth}
        \vspace*{4mm}
        \centering
        \includegraphics[width=1\linewidth]{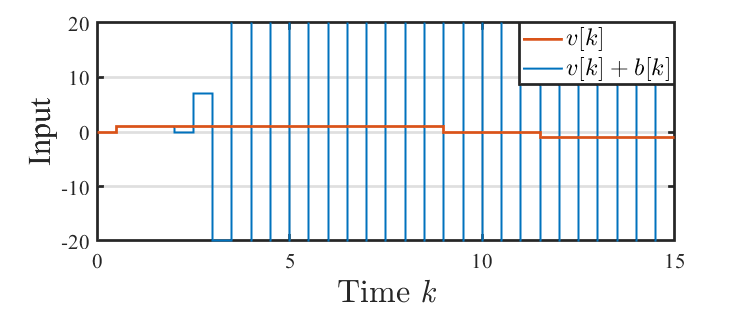}
        \subcaption{Control inputs and inputs contaminated by the attack.  \label{fig:no:dqzda}}
\vspace*{1mm}
    \end{minipage}
    \caption{Time responses of the system under the quantized zero dynamics attack based on
dynamic quantization
\label{fig:no:dqzdaresult}}
\end{figure}

(ii) Next, we look at attacks which are dynamically quantized from Section \ref{se:dqzda}. Fig.~\ref{fig:no:respnonmin1} presents the output of the plant similarly to the static quantization case in Fig.~\ref{fig:no:respnonminsta}. In Fig.~\ref{fig:no:diffy1}, the difference between the two outputs  with and without attacks is plotted, where the theoretical bound \eqref{eq:no:eps} from Theorem $1$ is $|CB|(1+|\lambda|){d}/{2}=0.038$. In comparison, it is impressive that the difference between the sampled output under attack is much closer to the normal output than in the case of static quantization, which shows the effectiveness of the approach. Fig.~\ref{fig:no:dqzda} displays the control input with and without attack (as in Fig.~\ref{fig:no:stazda}). Note that the attack effects do not appear for a while in the output. This is because the zero dynamics attack signal is initially small.

\begin{figure}[t]
    \begin{minipage}[h]{1\linewidth}
        \centering
        \includegraphics[width=1\linewidth]{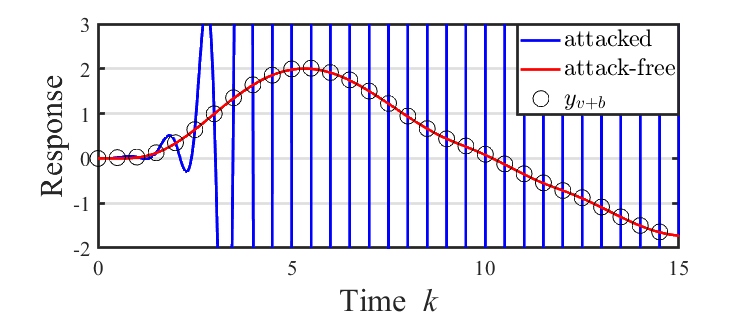}
        \subcaption{System responses  \label{fig:no:respnonmin2}}
    \end{minipage}\\
\vspace*{1mm}
    \begin{minipage}[h]{1\linewidth}
        \vspace*{4mm}
        \centering
        \includegraphics[width=1\linewidth]{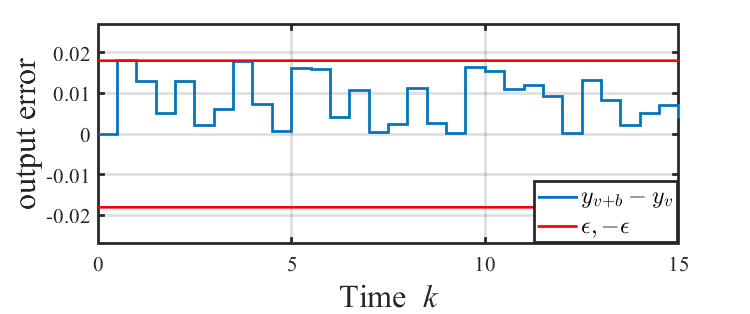}
        \subcaption{Difference in the outputs of the system  \label{fig:no:diffy2}}
    \end{minipage}\\
\vspace*{1mm}
    \begin{minipage}[h]{1\linewidth}
        \centering
        \vspace*{4mm}
        \includegraphics[width=1\linewidth]{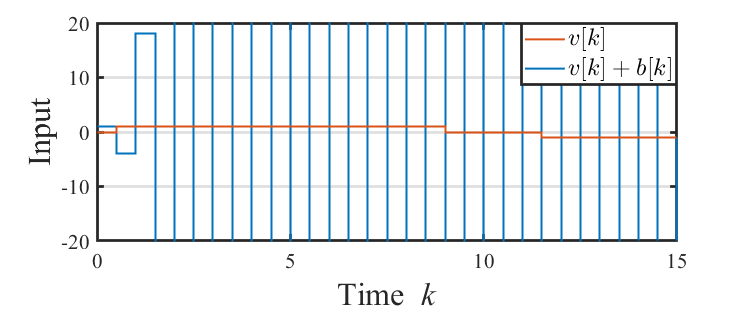}
        \subcaption{Control inputs and inputs contaminated by the attack.  \label{fig:no:qzdag}}
        \vspace*{1mm}
    \end{minipage}
    \caption{Time responses of the system under the attack via $\epsilon$-stealthy approach. \label{fig:no:qaresult}}
\end{figure}

(iii) Finally, we discuss the quantized attack based on $\epsilon$-stealthy approach from Section \ref{se:calc}.  The results are shown in Figs.~\ref{fig:no:respnonmin2}--\ref{fig:no:qzdag} as in the dynamic quantization case. Notice that the sampled output is closer to the attack-free one than in the previous two cases. In fact, the difference between attacked/original outputs remains below the bound $\epsilon=|CB|d+10^{-6}=0.018$ provided in Theorem \ref{theo:2}. Moreover, unlike the previous methods, in Fig.~\ref{fig:no:diffy2}, the attack effects appear at time $k=0$. This is because the initial value of the attack should be non-zero.

\section{Conclusion}\label{ch:conclusion}

In this paper, we have studied the effects of quantization on sampled-data 
systems under zero dynamics attacks. We have proposed methods for generating 
quantized attacks exploiting the unstable zeros in the plant after discretization. 
In particular, we have proposed two types of methods. The first one utilizes 
the conventional zero dynamics attacks and modifies them
via dynamic quantization. In the second method, slightly 
modified version of zero dynamics attacks is 
quantized. These methods have been compared with the simple
approach of directly quantizing the conventional zero
dynamics attacks, and our analysis shows that they perform
strictly better. 
In future research, we will study the effects of quantized attacks 
against discretized plants which are minimum phase as well as 
systems having feedback structures.

\bibliographystyle{plain}
\bibliography{main,BibDataBase_jun18a,BibDataBase_jun18}
\addtolength{\textheight}{-12cm}   






\end{document}